\documentclass{article}
\usepackage{hiph-art}
\usepackage{graphicx}
\volnumber{19} \issuenumber{1} \edyear{2004}                             
\frompage{000} \topage{000}                                              
\recrevdate{31 October 2005}                                              
\def\rightmark{Dynamical aspects of fragmentation}
\def\leftmark{M. J. Ison and C. O. Dorso}
 
\title{Dynamical aspects of fragmentation} 
\authors{ 
{M. J. Ison$^{1}$ and C. O. Dorso$^{1}$ %
\index{One, M. J. Ison} 
\index{Two, M. J. Ison} 
}\\[2.812mm]
{\normalsize
\hspace*{-8pt}$^1$ Departamento de F\'{\i}sica, Facultad de Ciencias Exactas y Naturales,
Universidad de Buenos Aires, Pabell\'on $I$, Ciudad Universitaria, Nu\~{n}ez, $1428$,\\
Buenos Aires, Argentina.\\[0.2ex] 
}}
 
\abstract{
In this short communication we address the problem of reducibility in a 
highly excited Lennard-Jones system. We show that the probability of 
emitting $n$ fragments can be described in terms of a single 
probability through the binomial expression. However, the Arrhenius law 
does not hold and the process can be viewed as a mixture of sequential 
and simultaneous fragmentation events.
}


\makeindex
\begin{document}
 
\maketitle

Fragmentation of hot nuclear systems has been the subject of several studies 
during the recent past. Because infinite nuclear matter has 
an equation of state very similar to that of a Van der Waals gas \cite{pandha}, 
which exhibits a liquid-gas phase transition, many generic approaches have been 
developed to provide a general framework for the nuclear problem including 
statistical models (like the lattice gas model) \cite{Das,richert} and dynamical
 ones (including classic dynamical models \cite{ison,dorso}).

The advantage of working with models which are fully microscopic is that both 
equilibrium and nonequilibrium features of the problem can be explored. 

Several aspects of the multifagmentation process are still a matter of debate. In 
particular, the problem of sequentiality versus simultaneity \cite{moretto,gross} arose since
Moretto and coworkers proposed that the complex behavior of fragment emission could 
be described in terms of a simple binomial distribution, namely:

\begin{equation}
P^m_n = \frac{m!}{n!(m-n)!} p^n (1-p)^{m-n}  \label{eqbinomial}
\end{equation}

Where $m$ stands for the number of "trials", and $n$ represent the number of successes.
Following \cite{moretto}, we associate the parameter $m$ to the maximum multiplicity 
for each energy and $p$ to the emission probability. It is interesting that $m$ could 
possibly be think of as the number of natural time intervals at which the system 
fragments with probability $p$. This approach rests on the assumption 
that a single fragmentation probability is capable of describing the emission process 
disregarding the characteristics of the emitting source. 

From the experimental side, it was found that in many reactions the multiplicity of 
intermediate-mass-fragments (IMF, with charges Z=3-20) is distributed as 
a function of the transverse energy $E_t$ binomially. This energy is assumed to be 
proportional to the excitation energy $E$, which is related to the temperature, considering the 
system as a Fermi gas, via $ T \propto \sqrt{E}$, so that $p \propto e^{-B/T}$, an Arrhenius law.       
Then, it is inferred that multifragmentation is a sequence of thermal binary events.
    
\begin{figure}[htbp]
\begin{center}
\setlength{\abovecaptionskip}{40pt} 
\includegraphics[totalheight=6cm]{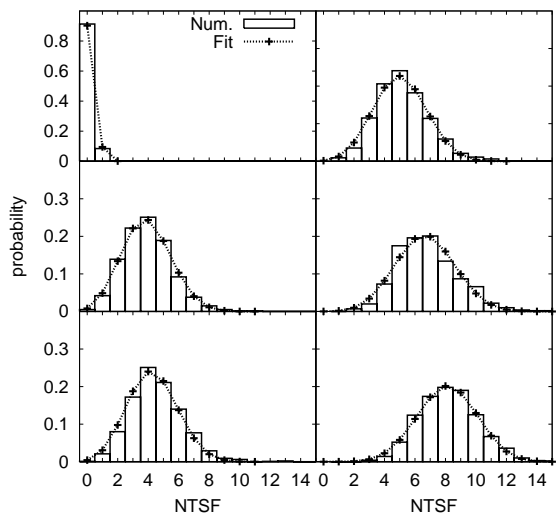} 
\end{center} 
\caption{Fragmentation probability as a function of the number of times the system 
fragments (NTSF) (histograms) and binomial fit (dotted line). For energies 
(from top to bottom and left to right): $E=-2.0\protect\epsilon, 
E=-0.2\protect\epsilon$, $E=0.0\protect\epsilon,
E=0.2\protect\epsilon$, $E=1.0\protect\epsilon$, and $E=2.0\protect\epsilon$}
\label{fig1}
\end{figure}

We performed molecular dynamics calculation of a system composed of $N=147$ particles interacting 
via a truncated Lennard-Jones potential. Initial configurations were prepared, like in previous 
works \cite{ison,dorso}, as hot dense drops. In order to get close to the experimental situation, no 
artificial constraining volumes were introduced, i.e. particles were freely evolving into vacuum.
 We define a fragmentation process when a source emits a quasi-stable fragment of at least 
four particles. Fragments are detected in configuration space and our temporal stability 
criterion is that a fragment is considered as quasi-stable when particles remain together for 
at least $5t_0$ (in natural Lennard-Jones units). 

In Figure \ref{fig1} we show that the distribution of emitted fragments is well adjusted 
with a binomial distribution. The agreement is remarkable specially when one 
considers that we are facing an out-of-equilibrium process and exploring a wide energy range, from 
$E=-2.0 \epsilon$, which has a U-shaped mass spectra, to $E=2.0 \epsilon$, characterized by an 
exponentially decaying mass distribution.  

\begin{figure}[htbp]
\begin{center}
\includegraphics[height=4cm,clip=]{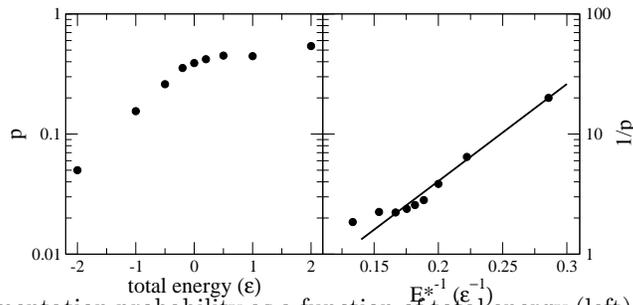} 
\end{center} 
\vspace*{-1.0cm}
\caption{Fragmentation probability as a function of total energy (left) and 
reciprocal of the probability as a function of the inverse of the excitation energy (right)} 
\label{fig2}
\end{figure}

To further analyze the implications of this approach we show in Fig.\ref{fig2} the relationship 
between the emission probability and the energy. It can be seen that a linear type 
behavior can only be identified for low energy values. To investigate if this relationship 
between $p$ and $E^*$ could indeed correspond to an Arrhenius law one would need to calculate 
the temperature of the emitting source. Is is easy to realize that one should have a 
dependence of the type $T \propto E^*$ in order to have an Arrhenius law. However, the 
temperature of the emitting source was calculated in Ref.\cite{ison} 
and the calculated values of $T$ were approximately linear with $E$ only for low energy values, 
 which suggest that the Arrhenius law could be fulfilled only when evaporation is the principal
 decaying mode.  

Moreover, an analysis of the fragmentation times was performed and showed that fragmentation 
events are more likely to occur at rather early stages of the evolution. The picture that 
emerges is that this process can be viewed as a mixture of sequential and simultaneous breakups.

\vfill\eject
\end{document}